\RequirePackage{fix-cm}
\documentclass[twocolumn]{svjour3}          
\smartqed  
\usepackage{amsmath,amssymb,amsfonts}
\usepackage{graphicx}
\usepackage{textcomp}
\usepackage{url}
\usepackage{makecell}

\usepackage{color}

\newcommand{\tabnotemark}[1]{$^{\mathrm{#1}}$}
\newcommand{\tablenote}[1]{\multicolumn{5}{l}{#1}\\}
\newcommand{\tablenotetwo}[1]{\multicolumn{4}{l}{#1}\\}

\begin{document}

\title{cuConv: A CUDA Implementation of Convolution for CNN Inference}


\author{Marc Jorda         \and
        Pedro Valero-Lara  \and
        Antonio J. Pe\~na
}


\institute{
              Barcelona Supercomputing Center (BSC) \\
              \email{marc.jorda,pedro.valero,antonio.pena@bsc.es}             \\
             \emph{This work has been submitted to the Springer for possible publication. Copyright may be transferred without notice, after which this version may no longer be accessible.}
}

\date{Received: date / Accepted: date}

\maketitle

\begin{abstract}
Convolutions are the core operation of deep learning applications based on Convolutional Neural Networks (CNNs). Current GPU architectures are highly efficient for training and deploying deep CNNs, and hence, these are largely used in production for this purpose. State--of--the--art implementations, however, present a lack of efficiency for some commonly used network configurations.

In this paper we propose a GPU-based implementation of the convolution operation for CNN inference that favors coalesced accesses, without requiring prior data transformations. Our experiments demonstrate that our proposal yields notable performance improvements in a range of common CNN forward propagation convolution configurations, with speedups of up to 2.29$\times$ with respect to the best implementation of convolution in cuDNN, hence covering a relevant region in currently existing approaches.
\end{abstract}

\section{Introduction}
Even though Convolutional Neural Networks (CNNs)\cite{cognitron,neocognitron,lecun98} were presented many years ago, since the appearance of CNNs in the
ImageNet Large Scale Visual Recognition Challenge (ILSVRC)
\cite{NIPS2012_4824_alexnet,ILSVRC15} in 2012, CNNs and other deep
neural network architectures have received considerable
more attention than in previous years. This interest comes from their impressive results in tasks
like image classification, speech recognition, and natural language
processing \cite{GuWKMSSLWW15_cnns}. Two main factors were necessary
to enable the success of CNNs: the availability of very large datasets
of manually annotated input data, and the high performance of current
computing systems. Large datasets are needed to train the deep neural
network parameters until a highly accurate result is reached. In turn,
such amount of data required the use of high performance computing
accelerators to keep the training time of deep neural networks within
reasonable limits. GPUs are probably the most common type of
accelerator used for deep learning, among others like FPGAs or custom solutions, and all of the main deep learning frameworks currently support execution on GPUs.

High performance computing is also important when CNN-based applications are deployed. Short response times are one of the most relevant parameters in terms of user satisfaction. Moreover, short latency requirements are mandatory for applications where delays in the response time pose safety implications. Deep learning-based safety-critical applications, like autonomous vehicles, will attain widespread use in the near future, which highlights the need for high performance CNN inference.

CNNs differ from other deep neural networks due to the inclusion of
\emph{convolutional layers}. 
The outputs of these layers are the result of weighted sums of inputs, like in fully-connected layers. 
The difference strives in the inputs that are involved in the computation of each
output, and the fact that convolutional layers feature a smaller
number of weights which are reused for several outputs.
Outputs of a convolutional layer are not a function of all the inputs as in fully-connected layers; they only depend on a contiguous subset of the inputs (known as receptive field).
This considerably reduces the computation
cost of these layers, when compared to fully-connected layers. It is also interesting that, in convolutional layers, computational and storage costs no longer depend on the number of inputs and outputs. The computational cost depends on the size of the contiguous subset of inputs, while the reuse of weights reduces the memory required to store them. 
To implement the weighted sums, convolutional layers use an operation called \emph{convolution}, which gives them the name.

Most of the execution time of a convolutional layer is spent performing convolutions. Since most of the layers in CNNs are convolutional, convolutions account for a large part of the overall network execution time. Several works target the optimization of convolution operations for GPUs. Some perform data transformations to be able to exploit already existing high-performance functions like GEMM \cite{Chetlur_cuDNN14}. Other works rely on algorithmic optimizations to reduce the computational cost of convolutions, thus improving their performance \cite{Lavin15b_winograd,VasilacheJMCPL14_fft}.

In this paper, we present the design and implementation of a GPU convolution algorithm for NVIDIA GPUs. Our approach is based on efficiently exploiting the GPU execution resources and in-core memories. Our design also does an efficient use of the GPU memory bandwidth, performing coalesced accesses without the need for costly data transformations before the main computation kernel. To evaluate our implementation of convolution for GPUs, we compare its performance with all the GPU convolution algorithms provided by the cuDNN library. This library is developed by NVIDIA and contains several implementations of convolution based on the current state--of--the--art algorithms. Currently, the convolutions and other deep learning operations provided by cuDNN are used as the GPU backend in most of the widely used deep learning frameworks, like PyTorch~\cite{pytorch} or TensorFlow~\cite{tensorflow}. For the experimental evaluation, we use the forward propagation convolution parameter configurations from five of the most well known CNN architectures, namely AlexNet~\cite{NIPS2012_4824_alexnet}, GoogleNet~\cite{SzegedyLJSRAEVR14_googlenet}, ResNet-50~\cite{HeZRS15_resnet}, SqueezeNet~\cite{IandolaMAHDK16_squeezenet}, and VGG19~\cite{SimonyanZ14a_vgg}. This ensures that our evaluation only focuses on relevant convolution configurations that are actually present in CNNs used in deep learning research and CNN-based applications. In total, we test more than 600 different convolution parameter configurations. As part of the evaluation, we also profile the execution of some representative parameter configurations to further explain the observed performance results.

This paper is organized as follows. In Section~\ref{sect_background} we introduce
the concepts required to understand this work. In
Section~\ref{sect_design} we present the design of our GPU convolution
implementation and discuss some of the trade-offs we faced during the
design. In Section~\ref{sect_eval} we describe the experimental
setup and present the obtained performance results. In Section
\ref{sect_relatedwork} we discuss other work related to convolution
optimizations. In Section \ref{sect_concl_future} we summarize this work's main conclusions and present possible future directions to improve our design.

\section{Background} \label{sect_background}

In this section we provide background information that will help readers understand the rest of the paper. First we present some details on convolutional layers and the convolution operation itself. Next, we introduce a few GPU architecture and programming model terms and naming conventions that will be used in the following sections. Finally, we briefly describe the three major state--of--the--art convolution algorithms, highlighting their strengths and weaknesses.

\subsection{Convolutional Layers}

\begin{figure}
    \centering
    \includegraphics[width=\columnwidth]{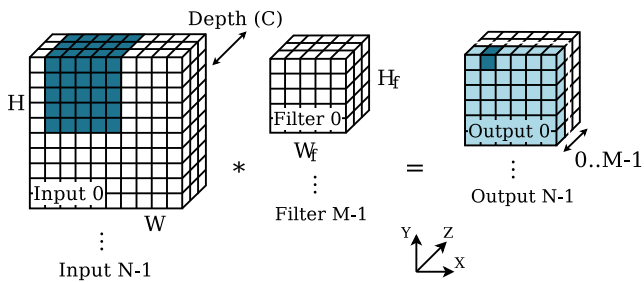}
    \caption{Convolution operations in a convolutional layer. The matrix produced by the convolution of Input 0 with Filter 0 is highlighted in light blue. The darker output element is the result of the dot product of Filter 0 with the highlighted subvolume of Input 0.}
    \label{conv_scheme}
\end{figure}

Convolutional layers in CNNs receive a set of \emph{N} 3D inputs (the \emph{input batch}) and generate an equally sized set of 3D outputs. The elements of each output volume are computed as a weighted sum of some of the elements of the corresponding input, which is usually transformed afterwards by a non-linear function (like sigmoid or ReLU~\cite{abs-1710-05941_activations}). These layers use convolution operations to perform the weighted sum of the input elements, hence their name. The weights of the layer are represented as a set of \emph{M} volumes called \emph{filters} or \emph{kernels}. Since the term \emph{kernel} is also used to describe a GPU function, we will use \emph{filters} to refer to the weights of convolutional layers. The convolution of a 3D input with a given filter produces a matrix. The matrices from the convolutions of a given input with all the layer's filters are stacked in the Z dimension to produce the corresponding output (see Figure~\ref{conv_scheme}). This process is repeated for all inputs to obtain the final set of \emph{N} output volumes.


A \emph{convolution operation} in the context of convolutional layers is a 2D discrete convolution, with the particularity that it uses a 3D input and filter. Each output element is computed as the dot product of the filter with a subvolume of the input, as depicted in Figure~\ref{conv_scheme}. The input subvolume used for each output element is different and depends on its coordinates. The top left element uses the subvolume in the top left corner of the input, and the rest of output elements are computed by translating the subvolume over the X and Y dimensions of the input (there is no translation over the Z dimension). The size of the input along the Z dimension (referred as the \emph{number of channels}) and the size of the filter along the Z dimension must be equal, for the dot products to be well defined. In this paper, we refer to the size in the Z dimension of inputs/filters/outputs as their \emph{depth}. The translation distance among two contiguous positions is called \emph{stride}, and defines the width and height of the output. Increasing the stride reduces the number of positions the filter can translate over the input, resulting in smaller outputs. \emph{Padding} is another parameter related to the output width and height. It defines the number of padding rows and columns added to the input along the X and Y dimensions, to increase its size. This increases the positions for the filter translation, increasing the output width and height. In CNNs, it is common to use a padding of $\frac{W_f-1}{2}$ elements in each side along the X dimension, and a padding of $\frac{H_f-1}{2}$ elements along the Y dimension. Combined with a stride of 1 element in both dimensions, these values generate an output with the same width and height of the input. 

Inputs, filters and outputs are all sets of 3D volumes. In CNNs, these sets are represented as 4D arrays usually called \emph{tensors}. In terms of the memory layout used to store the tensors, there is no standard agreement across the different deep learning frameworks. Several orderings of the four dimensions are being used, but two of the most common layouts are NCHW and CHWN \cite{mem_layout_Li16}. These letters stand for height (H), width (W), channels (C, another name for \emph{depth}), and N as the dimension representing the index of the volume within the set. The fourth dimension in the abbreviations is that with the elements contiguous in memory.

Most of the current deep learning frameworks use the operations from the cuDNN library as their GPU implementation of convolutions and other deep learning operations. cuDNN provides several implementations of the convolution operation, with different performance depending on the convolution parameters (input and filter sizes, stride, batch size, etc.). Because of this, several frameworks perform an initial exploration to choose the best-performing implementation of convolution for each convolutional layer.

\subsection{GPU Architecture and Naming Conventions}

\begin{figure}
    \centering
    \includegraphics[width=0.5\columnwidth]{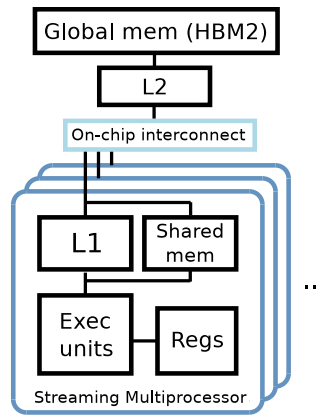}
    \caption{Schematic of our target GPU architecture.}
    \label{gpu_scheme}
\end{figure}

Throughout this paper we use NVIDIA terminology to refer to GPU architecture elements and to concepts from the GPU programming and execution models. A kernel is a function that contains the code that will be executed by each GPU thread. The number of GPU threads that will execute a kernel is specified when the kernel is launched, and is referred as the thread grid (from 1 to 3 dimensions). Threads are grouped into thread blocks. Threads in the same thread block are scheduled together in the same Streaming Multiprocessor (SM), and may be synchronized through thread barriers. Thread blocks are independent of each other and cannot synchronize their execution. The GPU execution model does not provide any guarantee on the thread blocks execution order; all possible interleavings of thread block execution are valid. This provides adaptive parallelism: a kernel launch with enough thread blocks may exploit the parallelism provided by GPUs with different number of SMs without requiring any modification from the developer.

In terms of the memory subsystem of our target GPU architecture, depicted in Figure~\ref{gpu_scheme}, we
assume a non-coherent 2-level cache hierarchy with a private L1 cache
per SM and an L2 cache shared by all the SMs. This configuration is
commonly found in discrete NVIDIA GPUs to connect the execution cores
to the on-board HBM2 memory. Besides the L1 cache, there are two other
types of in-core memories. Each SM features a software-managed
scratchpad memory, called \emph{shared memory} in CUDA
terminology. Each thread block running in an SM features its own space
within the shared memory. Only threads of the same thread block may
share data through shared memory. Accesses to shared memory are one or two orders of magnitude faster than accesses to global memory
(that miss\slash bypass the L1 cache) because these do not have to travel across the shared on-chip interconnect to reach the L2, nor reach global memory in case of an L2 miss. The last in-core memory is the register file, which stores the kernel's local variables, which are private to each thread. 

Exploiting in-core memories is one of the most important aspects to attain a high-performance GPU code, similar to a proper workload management to maximize cache hits in the CPU case. However, for GPUs, it is also important to consider the memory access pattern of contiguous threads. Execution units in GPU cores are vector units of a certain width, and a \emph{warp} describes the group of GPU threads that execute together as a single vector instruction. For memory loads and stores, threads in a warp may access any arbitrary address, but since these are executed in lock-step, all the threads will be stalled until all the memory transactions are fulfilled. Thus, to attain the minimum memory instruction cost, all threads in a warp have to access addresses within the same cache line (usually referred as \emph{coalesced} accesses). Current NVIDIA GPUs have a warp width of 32 threads, and an L1 cache line size of 128 bytes.

\subsection{Convolution Algorithms}

There are various algorithms that may be used to compute a convolution. The first option is to directly apply the convolution formula, but there are more complex approaches.
The convolution inputs and filters can be transformed to be able to exploit high-performance matrix-matrix multiply (GEMM) implementations~\cite{Chetlur_cuDNN14}.
Winograd's minimal filtering algorithm~\cite{Lavin15b_winograd} and Fast Fourier Transform (FFT)~\cite{MathieuHL13_fft,VasilacheJMCPL14_fft} are two other approaches to compute convolutions, both based on reducing the number of multiplications (but increasing the number of additions) to reduce the computational cost of the computation.

\subsubsection{GEMM}

The GEMM-based algorithm transforms the filters into an intermediate matrix, where each row contains the elements of a given filter, flattened as a 1D array. The inputs are also transformed to an intermediate matrix, in such a way that the dot products computed by the GEMM of these two matrices match the ones that would be computed by the convolution formula. Finally, the GEMM's output matrix is transformed to obtain the convolution output with the appropriate format. 
GEMM-based algorithms may benefit from already existing high-performance GEMM on GPUs \cite{DongarraICCS17}. However, the
intermediate matrices require a significant amount of memory. Specifically, the one derived from the convolution inputs has to store duplicate elements due to the overlap of the filter positions when the stride is smaller than the filter, which require more memory than the inputs themselves. To avoid this memory cost, some variants of the GEMM based convolution algorithms do not explicitly store the intermediate matrices and do the transformation when loading the data in the in-core memories (e.g. shared memory, registers). 

\subsubsection{Winograd}

The Winograd-based algorithm applies the transformations described in \cite{winograd1980arithmetic} to minimize the number of multiplications needed for the computation of the convolution. This number is reduced by increasing the number of additions, which are usually less costly to compute. The number of additions increases quadratically with the size of the input, making this technique better suited for small inputs. For large inputs, the Winograd-based algorithm divides the input in tiles of $p\times{}p$ elements, which have to overlap by $p-2$ elements. After computing the minimal convolution for each tile, the intermediate results are combined to obtain the final convolution output.

\subsubsection{FFT}

The FFT-based convolution algorithms exploit the property that the convolution in the time domain is equal to point-wise multiplication in the Fourier (frequency) domain. 
The algorithm computes the FFT of the convolution inputs, then performs the point-wise multiplication followed by an inverse FFT to get the convolution output.
While these transformations may be too costly to compute a single convolution, their cost can be amortized by several convolutions. This is the case of convolutional layers where the FFT of each input and filter can be reused for all the convolutions where these participate. Thus, the potential improvement of FFT-based algorithms increases with larger number of inputs and\slash or larger number of filters.

\section{Design and Implementation} \label{sect_design}


In this section we present the details of the design of our GPU-based algorithm to compute convolution. We also discuss and pay particular attention to some relevant implementation trade-offs.

\begin{figure}
    \centering
    \includegraphics[width=0.75\columnwidth]{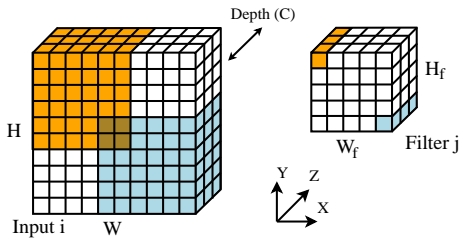}
    \caption{
    Reuse of input data for two example rows of a filter (highlighted in blue and orange), for a convolution with a stride of 1. The highlighted input rows (two sets of 6x6 rows) are point-wise multiplied with the filter row of the same color during the convolution computation.}
    \label{conv_reuse}
\end{figure}

As introduced earlier, our efforts focus on efficiently exploiting the GPU hardware features to implement a fast GPU implementation for convolutional operations. To implement a fast GPU kernel, an effective exploitation of the top-level hierarchies of memory (shared memory and registers) is of vital importance, in order to maximize the reuse of data. This is in fact particularly necessary for those throughput-oriented architectures, such as GPUs, where the accesses to memory are impacted by a high latency.
In the computation performed by a convolutional layer we find two levels of data reuse:
\begin{enumerate}
\item \textbf{At the layer level.} Convolutional layers receive a set of inputs that have to be convolved with a set of filters. At this level we may exploit the reuse of data in both, the filters and the inputs. Each input is convolved with all the filters or, the other way around, each filter is convolved with all the inputs. 
\item \textbf{At the convolution level.} Inside a given convolution for a specific pair of input and filter, the filter elements (i.e. weights) are shared by several elements of the same input, which is one of the main differences when compared to the fully-connected layers.
As the filter is passed (computed) through all the elements of the input, each filter row along the Z dimension is point-wise multiplied with the corresponding rows along the Z dimension of the input (to make the presentation less verbose, for the rest of this section we will refer to the rows along the Z axis simply as \emph{rows}). Besides, most of the convolutional operations to be performed present a stride of 1 in both dimensions, X (horizontal) and Y (vertical), which helps to a higher data reuse, since the input elements which share one row of a given filter generate a cube composed by contiguous rows (memory positions), as depicted in Figure~\ref{conv_reuse}. This scenario is particularly well-suited for GPU computing.
\end{enumerate}

Since the filter size is usually smaller than the input size, it is better to exploit the reuse of data from the filter side than from the input side, because it requires less shared  memory and each value brought to the shared memory will be reused more times.  
It is important to note that the reuse of data for those rows closer to the edges involves less filter-input operations. This is particularly negative in the rows positioned in the corners, which may be only used once. 
On the other extreme, those rows positioned around the center of the input (with coordinates $W_f-1 \leq x \leq W-W_f$ and $H_f-1 \leq y \leq H-H_f$) present the best ratio of data reuse ($H_f \cdot W_f$ times per element). 
The data reuse of any row of a particular filter is equal to $(W-W_f+1) \cdot (H-H_f+1)$, which is the product of the height and width of the outputs of the convolution.

\begin{figure}
    \centering
    \includegraphics[width=0.75\columnwidth]{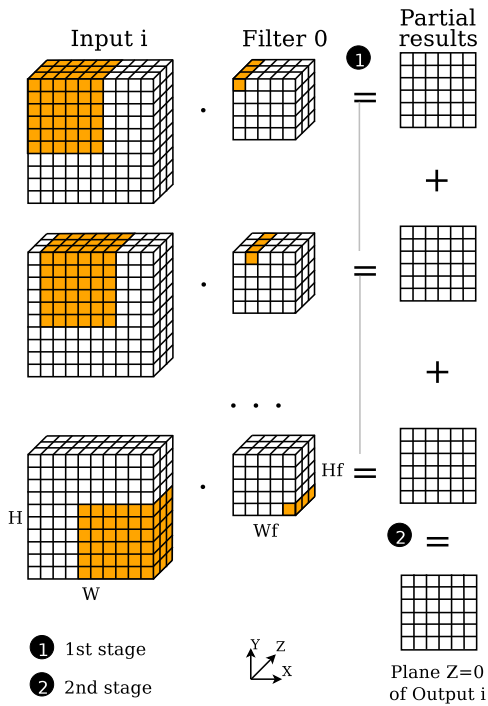}
    \caption{Stages of our implementation of convolution depicted for an arbitrary input and the first filter of the convolutional layer. Scalar products in the first stage generate partial results which are aggregated in the second stage to obtain the final output elements.}
    \label{conv_our_impl}
\end{figure}

Our GPU convolution is divided into two stages, implemented in two separate kernels. In the first stage, we compute all the dot products between input and filter rows that are needed by all the convolutions of the convolutional layer. The result of this first step is a set of $H_f \cdot W_f \cdot N \cdot M$ temporary matrices, each of size $H-H_f+1$ by $W-W_f+1$, as depicted by the scheme from Figure~\ref{conv_our_impl}. In other words, the first kernel applies a reduction to the contiguous input 3D patches along the Z dimension for all filters and inputs, resulting in a collection of temporary 2D matrices with the size of an output X-Y plane.

The second stage collects and sums the $H_f \cdot W_f$ temporary matrices resulting from a given input-filter pair, and stores the result as the corresponding output plane. In Figure~\ref{conv_our_impl}, the output plane for the second stage is the result of adding the matrices of partial results depicted in the scheme. Output elements are the sum one element from each 2D matrix (i.e. filter row), which is one of the reasons why this final reduction is performed in a separate kernel.

The division into two stages enables us to exploit the reuse of filter elements during the first stage. Each thread block is responsible for the dot products of one (possibly more) filter row with all the rows from the inputs that interact with it in any of the convolutions. In order to leverage the in-core memory regions, the threads of a given thread block collaborate to bring the filter row into a shared memory buffer. Once the filter data is ready, each thread computes one (or several) of the dot products involving the assigned filter row, and stores the result in a temporary output array. 

Depending on the input size and the number of inputs and filters, the dot products of one filter row may be split into several thread blocks. This may be necessary when the number of threads per block required to perform all the dot products is larger than the maximum number of threads per block allowed by the GPU. It can also be a possible optimization to increase the thread block-level parallelism and exploit all the SMs in configurations that would otherwise feature a low number of thread blocks. However, this increases the overall amount of long-latency memory accesses because the kernel has to move the same filter row data several times from global memory to shared memory, which may overcome the possible benefit of increasing the thread block-level parallelism (the GPU L2 cache may reduce this impact depending on workload size and the temporal locality of the thread blocks which access the same filter row).

In terms of data layout, our implementation supports tensors in NCHW
format, which is one of the most common memory layouts in CNNs
\cite{mem_layout_Li16}. In this format, elements in the X dimension
are stored in contiguous memory locations. Then it stores the rows
along Y, the planes along Z, and finally the rest of the 3D
inputs\slash filter\slash outputs as the fourth dimension. The
advantage of this format, given the fact that the input rows that
reuse a given filter row in a stride 1 convolution are contiguous in
the X dimension, is that the reads of the input data performed by the
first-stage kernel are almost perfectly coalesced. Unlike other
implementations, like the GEMM-based
convolution~\cite{Chetlur_cuDNN14}, the coalesced accesses are
attained without the need for data-layout transformations of the inputs
to adapt their layout to the memory access pattern of the kernel. For
convolution configurations where the width of the contiguous input 3D
patch read by a thread block is multiple of 32, all the warps will
perform coalesced accesses. For other configurations, some of the
warps will read elements from two different rows along the X axis of the 3D patch,
which introduces a gap of $W_f-1$ elements in between the memory positions accessed by two contiguous threads (see Figure~\ref{conv_reuse}). In such a case, the memory load instruction will issue at least 2 memory accesses, instead of the ideal single access. However, since this only occurs to some of the warps, the cost of transforming the input memory layout to always feature perfectly coalesced reads would overcome the benefits obtained by avoiding the additional memory accesses.

For convolutions which involve filters of size 1$\times$1, the second kernel is not necessary because each output element only depends on one input and one filter row, and thus the outputs of the first kernel are already the final output elements.
In this case, the first-stage kernel stores its outputs with the proper memory layout to represent the final 4D output tensor, avoiding the need to execute the second-stage kernel.

\section{Experimental Evaluation} \label{sect_eval}


Following the study of~\cite{access}, to analyze the performance of
our proposed implementation of convolution we selected all the
forward propagation convolutional layer configurations from five widely known CNNs:
AlexNet~\cite{NIPS2012_4824_alexnet},
GoogleNet~\cite{SzegedyLJSRAEVR14_googlenet},
ResNet-50~\cite{HeZRS15_resnet},
SqueezeNet~\cite{IandolaMAHDK16_squeezenet}, and
VGG19~\cite{SimonyanZ14a_vgg}, identifying the input and filter sizes,
the number of filters, and the depth (see Table~\ref{tab_cnns}). For the number of inputs (batch size), we evaluate a range of characteristic values (1, 8, 16, 32, 64, 128, 256) because this parameter is usually selected by the end user of the CNN. 
The final set includes configurations with the following filter sizes: 1$\times$1, 3$\times$3, and 5$\times$5. The most common filter size is 1$\times$1, which features a particular purpose within CNNs. Convolutional layers with this filter size are used as a means to control the computational cost of the following convolutional layer, which usually has a larger filter size. Tuning the number of filters in the 1$\times$1 layer, to reduce the depth of the outputs, directly translates in less computational load in the following layer.

\begin{table*}
\scriptsize
\caption{Summary of features of the convolution operations (with
  stride 1) present in the five well-known CNNs we selected for our
  evaluation~\cite{access}.}
\label{tab_cnns}
\begin{tabular*}{\textwidth}{@{}@{\extracolsep{\fill}}lccccc@{}}\text

 & \multicolumn{1}{c}{\textbf{GoogleNet}} & \multicolumn{1}{c}{\textbf{SqueezeNet}} & \multicolumn{1}{c}{\textbf{AlexNet}} & \multicolumn{1}{c}{\textbf{Resnet50}} & \multicolumn{1}{c}{\textbf{VGG19}} \\\hline
\textbf{\makecell[l]{\# Distinct convolution \\ configurations}} & 42 & 21 & 4 & 12 & 9 \\\hline
\textbf{Network input size} & \multicolumn{1}{c}{224$\times$224$\times$3} & \multicolumn{1}{c}{224$\times$224$\times$3} & \multicolumn{1}{c}{224$\times$224$\times$3} & \multicolumn{1}{c}{224$\times$224$\times$3} & \multicolumn{1}{c}{224$\times$224$\times$3} \\\hline
\textbf{\makecell[l]{Input size to last \\ convolutional layer}} & \multicolumn{1}{c}{7$\times$7$\times$832} & \multicolumn{1}{c}{13$\times$13$\times$512} & \multicolumn{1}{c}{13$\times$13$\times$384} & \multicolumn{1}{c}{7$\times$7$\times$1024} & \multicolumn{1}{c}{14$\times$14$\times$512} \\\hline
\textbf{\makecell[l]{Convolution filters sizes \\ (\% of conv. configs.)}} & \multicolumn{1}{c}{1$\times$1 (57.2\%)} & \multicolumn{1}{c}{1$\times$1 (71.4\%)} & \multicolumn{1}{c}{3$\times$3 (75\%)} & \multicolumn{1}{c}{1$\times$1 (66.7\%)} & \multicolumn{1}{c}{3$\times$3 (100\%)} \\
 & \multicolumn{1}{c}{3$\times$3 (23.8\%)} & \multicolumn{1}{c}{3$\times$3 (28.6\%)} & \multicolumn{1}{c}{5$\times$5 (25\%)} & \multicolumn{1}{c}{3$\times$3 (33.3\%)} &  \\
 & \multicolumn{1}{c}{5$\times$5 (19\%)} &  &  &  &  \\

\hline
\end{tabular*}
\end{table*}

Like the filter sizes, we test configurations with different input sizes and number of filters. The input size (H and W; all of the five selected CNNs use square inputs and filters) ranges from 7 to 224, while the number of filters varies between 16 and 2,048. All the configurations implement a stride of 1 element and an input padding of $\frac{W_f-1}{2}$ and $\frac{H_f-1}{2}$ for the X and Y dimensions, respectively. Our experiments were performed with data represented as single-precision floating-point values. 

We compare our implementation of convolution for GPUs with those implementations available in the NVIDIA CUDA Deep Neural Network library (cuDNN). Among other operations used in deep neural networks, cuDNN offers several implementations of convolution based on state--of--the--art algorithms (GEMM, FFT, and Winograd). Most of the well-known deep learning frameworks (like TensorFlow, PyTorch, Theano, etc.) rely on cuDNN to execute operations on NVIDIA GPUs.

cuDNN provides several variants of each of the mentioned convolution algorithms. See Table~\ref{cudnn_algos} for a detailed list with a short description of their differences. However, since cuDNN is a closed-source library, it is difficult to assess the exact details of the implementation of its convolutions. The latest cuDNN version provides a helper function that uses heuristics to suggest the fastest convolution variant for a specific convolution parameters configuration. However, there is not much information on which heuristics are used, and the suggested variant is not guaranteed to be the fastest. Thus, we run all the convolution variants for all the configurations, to ensure that we are comparing our results against the lowest cuDNN time.

\begin{table*}
\caption{Convolution algorithm variants available in cuDNN.}
\label{cudnn_algos}
\scriptsize
\begin{tabular*}{\textwidth}{@{}@{\extracolsep{\fill}}lll@{}}
\hline

\textbf{Algorithm} & \textbf{Variant} & \textbf{Description} \\\hline
GEMM & --- & \multicolumn{1}{l}{The transformed input matrix is explicitly generated before} \\
     &     & \multicolumn{1}{l}{the GEMM kernel} \\
     & Implicit & \multicolumn{1}{l}{The input transformation is performed on--the--fly by the kernel that} \\
     &          & \multicolumn{1}{l}{computes the GEMM} \\
     & Implicit precomp. & \multicolumn{1}{l}{Like \emph{Implicit}, but another kernel precomputes offsets used in the} \\
     &                   & \multicolumn{1}{l}{implicit transformation} \\
\hline
\rule{0pt}{2.5ex}FFT & --- & Baseline FFT-based convolution  \\
 & Tiled & \multicolumn{1}{l}{The inputs are processed in tiles to reduce the temporary storage} \\
 &       & \multicolumn{1}{l}{required} \\
\hline
\rule{0pt}{2.5ex}Winograd & --- & A single kernel performs the Winograd transforms and multiplication  \\
 & Non-fused & \multicolumn{1}{l}{The Winograd transform of inputs, filters and outputs is performed} \\
 &           & \multicolumn{1}{l}{in separate kernels} \\

\hline
\end{tabular*}
\end{table*}


The convolution algorithms in cuDNN experience some parameter limitations (which are different for each algorithm), yielding some of the implementations unavailable for certain convolution configurations. Like our implementation, most of the cuDNN convolution variants require an additional buffer in GPU memory to store intermediate results, with varying size depending on the algorithm and the convolution parameters. We limit the temporary allocation size to 1~GB. This only affects around 4\% of algorithm/configuration cases, and most of these would not be relevant even without the space limitation, because these show poor performance compared to other algorithms for the configurations where these require more than 1~GB of additional space.

Our test platform is an IBM POWER9 server running Red Hat Enterprise Linux Server 7.4. The GPU installed in this system is an NVIDIA Tesla V100-SXM2 (Volta generation), and the GPU software stack available is CUDA version 9.2 with cuDNN version 7.1.

\subsection{Speedup over cuDNN}

\begin{figure*}[!t]
    \centering
    \includegraphics[width=\textwidth]{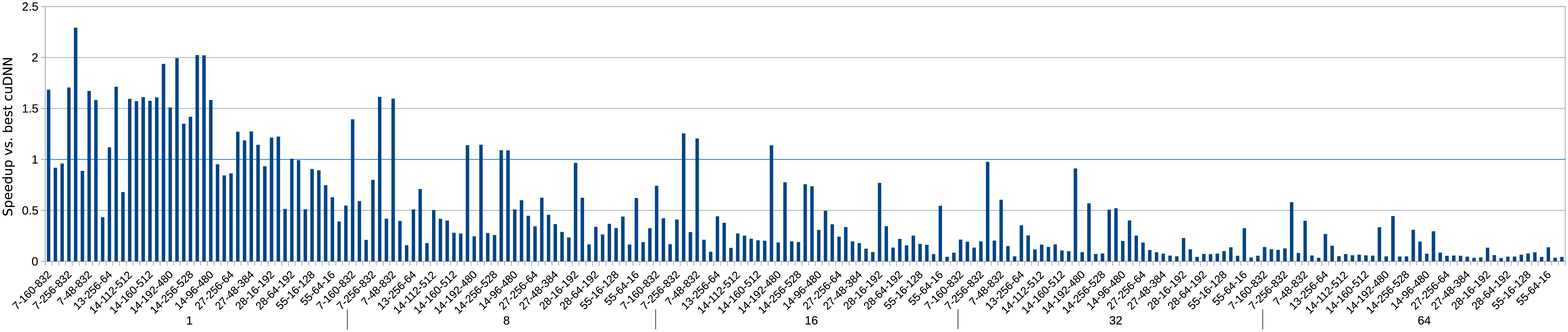}
    \caption{Speedup of our implementation of convolution w.r.t. the best performing cuDNN algorithm for each configuration. Configurations with $1\times1$ filters and batch size up to 64. Labels are formatted as [inputs X\&Y size]-[number of filters]-[depth].}
    \label{speedup_1x1}
\end{figure*}

\begin{figure*}[!t]
    \centering
    \includegraphics[width=\textwidth]{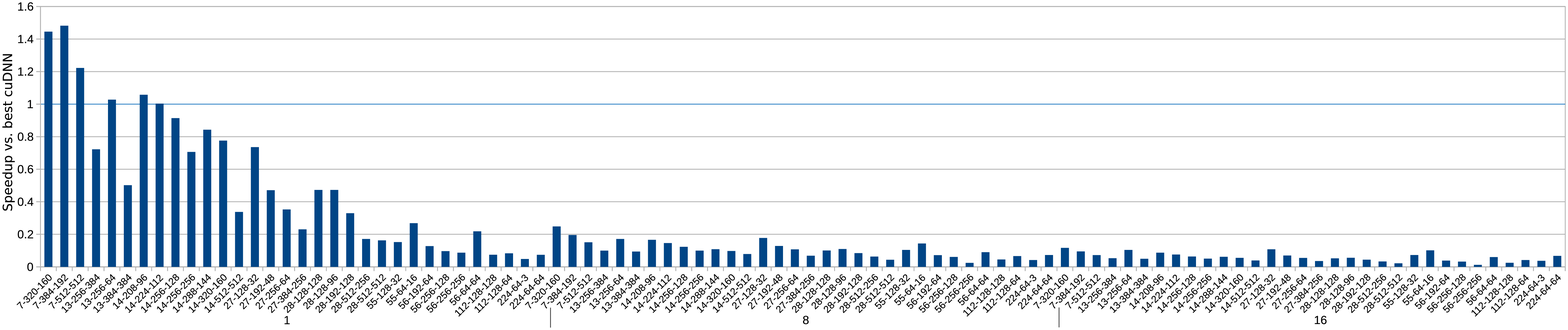}
    \caption{Speedup of our implementation of convolution w.r.t. the best performing cuDNN algorithm for each configuration. Configurations with $3\times3$ filters and batch size up to 16. Labels are formatted as [inputs X\&Y size]-[number of filters]-[depth].}
    \label{speedup_3x3}
\end{figure*}

\begin{figure*}[!t]
    \centering
    \includegraphics[width=\textwidth]{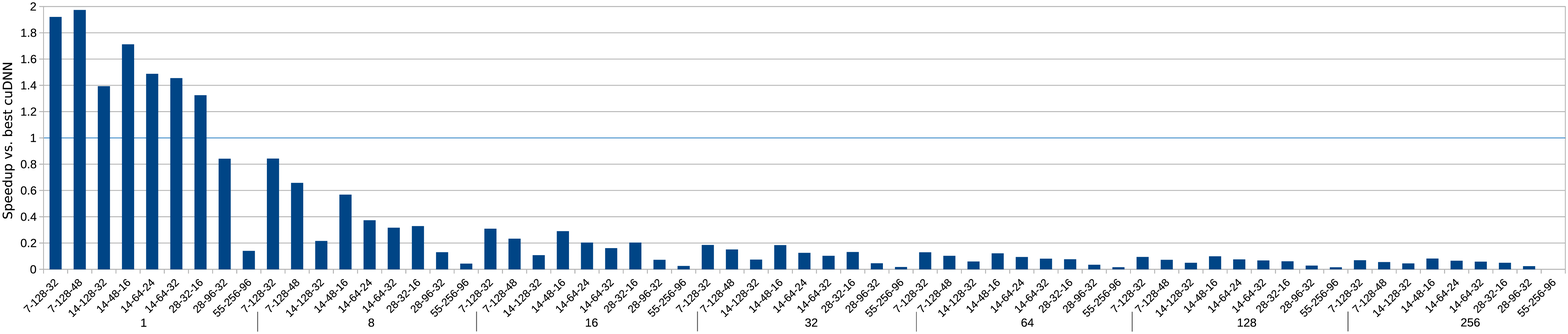}
    \caption{Speedup of our implementation of convolution w.r.t. the best performing cuDNN algorithm for each configuration. Configurations with $5\times5$ filters and batch size up to 256. Labels are formatted as [inputs X\&Y size]-[number of filters]-[depth].}
    \label{speedup_5x5}
\end{figure*}

We start by comparing the performance of our implementation with the
best time of all the convolution algorithms provided by cuDNN. For the
sake of clarity, given the number of tested configurations, we present
the speedup results in three different plots, separating the
configurations by filter size (figures~\ref{speedup_1x1},
\ref{speedup_3x3} and \ref{speedup_5x5}). We will keep this format in
the rest of this section, especially when analyzing the reasons behind
the observed performance. In fact, the filter size is the most
influential parameter and determines the best performing cuDNN algorithm for a given configuration \cite{access}. Also, even though all configurations were tested with batch sizes up to 256, figures \ref{speedup_1x1} and \ref{speedup_3x3} show less batch sizes to focus on the relevant results. Note also that the configuration parameters within a given batch size (input size, number of filters and depth) are different in each of the plots because these depend on the convolution configurations present in the selected CNNs. All the reported results are the mean of nine different executions.

\begin{table*} 
\caption{Kernel execution time ($\mu$s) for the selected configurations with 1$\times$1 filters.}
\label{times_1x1}
\scriptsize
\begin{tabular*}{\textwidth}{@{}@{\extracolsep{\fill}}llrrr@{}}
\hline

\textbf{Algorithm} & \textbf{GPU kernel} & \multicolumn{1}{c}{\textbf{A}} & \multicolumn{1}{c}{\textbf{B}} & \multicolumn{1}{c}{\textbf{C}} \\\hline
\textbf{GEMM} & implicit\_convolve\_sgemm\textless{}1024\textgreater{}\tabnotemark{a} &  &  & 19.20 \\
\textbf{implicit} & implicit\_convolve\_sgemm\textless{}128\textgreater{}\tabnotemark{a} & 128.13 & 47.87 &  \\

 & \textit{\textbf{Total}} & \textit{128.13} & \textit{47.87} & \textit{\textbf{19.20}} \\
 &  &  &  &  \\

\textbf{GEMM} & computeOffsetsKernel & 1.98 & 2.00 & 1.89 \\
\textbf{implicit} & volta\_scudnn\_128x64\_relu\_interior & 105.31 & 43.23 & 22.40 \\

\textbf{precomp.} & \textit{\textbf{Total}} & \textit{107.29} & \textit{\textbf{45.23}} & 24.29 \\
 &  &  &  &  \\
\textbf{Our} & scalar\_prods\_kernel & 58.56 & 73.86 & 22.53 \\
\textbf{impl.} & \textit{\textbf{Total}} & \textit{\textbf{58.56}} & \textit{73.86} & \textit{22.53} \\


\hline
\tablenote{\rule{0pt}{2.5ex}\textbf{A}: 7-1-1-256-832, \textbf{B}: 14-1-1-1024-256, \textbf{C}: 27-1-1-256-64.}
\tablenote{\tabnotemark{a}Some kernels show different versions depending on the configuration.}
\end{tabular*}
\end{table*} 

All three plots show a similar trend, where our implementation shows a clear advantage over the best cuDNN variant for batch size of 1, although this advantage is reduced as the batch size and the input X and Y size increase. For configurations with 1$\times$1 filters and batch size 1, the average speedup obtained by our implementation is 1.23$\times$, with a maximum of 2.29$\times$ for the 7-32-832 configuration (the numbers correspond to [input size in X and Y]-[number of filters]-[depth]). Convolutions with 1$\times$1 filters are 52.3\% of the tested configurations, the most common case in the dataset.

The configurations with 3$\times$3 filters are those where our
implementation is less competitive. Except for the three leftmost
configurations, the performance attained is equal or lower than the
best cuDNN variants (mainly the two based on Winograd). The Winograd
convolution was originally published for 3$\times$3 filters
\cite{Lavin15b_winograd} and obtains notably high performance compared to the other cuDNN variants in the tested configurations with this filter size; in around 40\% of the cases the second highest performing variant is at least 50\% slower than one of the two Winograd variants.

For 5$\times$5 filters, the results are similar to the 1$\times$1
filter case. Our implementation shows a notable advantage for batch
size of 1, with an average speedup of 1.36$\times$ and a maximum of
1.97$\times$ over the fastest cuDNN variant. Winograd is still the
fastest cuDNN algorithm for most configurations with this filter size,
although experiencing a smaller advantage compared to the 3$\times$3 cases.

Overall, our proposed implementation is faster than the best cuDNN
variant in 8.31\% of the tested configurations, with an average speedup of 1.46$\times$ for these 50 configurations. Almost all of them have a batch size of 1, which makes our implementation especially suited for the inference stage of CNN-based applications. Given that most of the machine learning frameworks automatically select the best-performing convolution algorithm for each convolutional layer, our implementation will improve the performance of layers with such configurations, without affecting the performance of the rest.

\subsection{Algorithms Performance Comparison}

To further analyze the behavior of our implementation, we profiled the execution of a set of representative configurations to see the kernels involved in each analyzed algorithm and their features.

Table~\ref{times_1x1} contains the kernel execution times for three selected configurations with 1$\times$1 filters. Besides our implementation's time, we include the two implicit GEMMs because these are the cuDNN variants that provide the highest performance for most of the cases with this filter size~\cite{access}. GEMM-impl-precomp is the only presented variant with more than one kernel; however, the \emph{volta\_scudnn\_128x64\_relu\_interior} kernel clearly dominates the total execution time.
Configuration \textbf{A} is one of the configurations where our implementation shows a notable advantage over cuDNN. 
\textbf{B} (twelfth bar from the left in Figure~\ref{speedup_1x1}) and
\textbf{C} are examples of  configurations where the cuDNN variants
are faster than our implementation. From the profiling we observe that
our kernel is launched with a larger number of thread blocks compared
to the GEMM variants. For \textbf{A}, we launch 256 thread blocks,
while GEMM-impl and GEMM-impl-precomp launch 16 and 4,
respectively. Thus, we expose more thread block-level parallelism,
which gives our implementation a clear advantage. However, our
approach is slower than the GEMM variants for configuration
\textbf{B}, where we launch 1,024 thread blocks, GEMM-impl 224 and
GEMM-impl-precomp 32. In this case, the overhead associated with
thread block setup and termination affects negatively our
implementation, and hence the GEMM variants perform higher.
Configuration \textbf{C} features the same number of filters than
\textbf{A}, but the inputs size is larger, with almost 15 times more elements per input. Our implementation's time is close to both GEMM variants, but its performance is more impacted by the input size increase. Since we increase the number of threads per block to cope with a larger input size, a work-fusion optimization at the thread level could be explored as a possible optimization (see Section~\ref{sect_concl_future} for further discussion).

Kernel times for two configurations with 3$\times$3 filters are presented in Table~\ref{times_3x3}. 
In this case, we include the two Winograd variants because those are
the highest performing cuDNN algorithm in most configurations with
this filter size. We keep the GEMM-implicit-precomp times to show the
notable advantage of Winograd even within the cuDNN convolution
implementations. Compared to GEMM-impl-precomp, Winograd non-fused is
1.93$\times$ faster for configuration \textbf{A}, and Winograd is
1.67$\times$ faster for configuration \textbf{B}. 

All the variants shown in Table~\ref{times_3x3} are composed by
multiple kernels, including our implementation, which features kernels
corresponding to each of the two stages as detailed in Section \ref{sect_design} (in Table~\ref{times_1x1} the second kernel does not appear because it is not needed for configurations with 1$\times$1 filters). Comparing the times of the two kernels, we observe that the overall time of our implementation is dominated by the first stage. For the configurations shown in Table~\ref{times_3x3}, the second stage is 8.5\% of the total time for \textbf{A} and only 1.14\% of the total time for \textbf{B}.

Our implementation is clearly the fastest for configuration
\textbf{A}. This is because of our better exploitation of thread-block
parallelism for configurations with small inputs. Nevertheless, our implementation is also the most affected by the increase of input size and depth from configuration \textbf{B}. The profiling obtained for \emph{scalar\_prods\_kernel} shows 
that the compute and memory units utilization is around 50\%, even though the kernel shows an occupancy of 86\%. This further indicates an improvement direction, as we discuss later in Section~\ref{sect_concl_future}.

\begin{table} 
\caption{Kernel execution time ($\mu$s) for the selected configurations with 3$\times$3 filters.}
\label{times_3x3}
\scriptsize
\begin{tabular*}{\columnwidth}{@{}@{\extracolsep{\fill}}llrr@{}}
\hline
\textbf{Algorithm} & \textbf{GPU kernel} & \multicolumn{1}{c}{\textbf{A}} & \multicolumn{1}{c}{\textbf{B}} \\\hline
\textbf{Winograd} & generateWinogradTilesKernel & 9.12 & 19.77 \\
 & winograd3x3Kernel\textless{}1,4,8\textgreater{} & 101.91 & 212.58 \\
 & \textit{\textbf{Total}} & \textit{111.03} & \textit{\textbf{232.35}} \\
 &  &  &  \\
\textbf{Winograd} & winogradForwardData4x4 & 8.06 & 22.75 \\
\textbf{non-fused} & winogradForwardFilter4x4 & 17.44 & 35.10 \\
 & volta\_sgemm\_128x64\_nn & 69.31 & 242.56 \\
 & winogradForwardOutput4x4 & 10.82 & 27.14 \\
 & \textit{\textbf{Total}} & \textit{105.63} & \textit{327.55} \\
 &  &  &  \\
\textbf{GEMM} & computeOffsetsKernel & 1.98 & 2.11 \\
\textbf{implicit} & volta\_scudnn\_128x64\_relu\_interior & 201.47 & 386.97 \\
\textbf{precomp.} & \textit{\textbf{Total}} & \textit{203.45} & \textit{389.08} \\
 &  &  &  \\
\textbf{Our} & scalar\_prods\_kernel\_simple & 52.86 & 461.37 \\
\textbf{impl.} & sum\_kernel\_simple & 4.93 & 5.31 \\
 & \textit{\textbf{Total}} & \textit{\textbf{57.79}} & \textit{466.68} \\

\hline
\tablenotetwo{\rule{0pt}{2.5ex}\textbf{A}: 7-1-3-384-192, \textbf{B}: 13-1-3-384-384.}
\end{tabular*}
\end{table} 

For configurations with 5$\times$5 filters, we selected two
configurations with different batch size to see their effect on the
algorithm performance. The execution time of the Winograd kernels in
configuration \textbf{A} (Table~\ref{times_5x5}) shows that the pre-
and post-processing kernels impacts significantly the execution time
of convolutions with relatively small computational load. In this specific case, the sum of the \emph{winogradForward...4x4} kernels is 53.3\% of the total Winograd time. Even if we consider only the \emph{volta\_sgemm\_128x64\_nn} time, our implementation's time is still faster, following the trend for 1$\times$1 and 3$\times$3 filters. Comparing the times with configuration \textbf{B}, we observe that Winograd scales better with the batch size than our implementation, as it is expected from arithmetic strength reduction optimization approaches, which are more suited for convolution configurations with larger computational loads.

\begin{table} 
\caption{Kernel execution time ($\mu$s) for the selected configurations with 5$\times$5 filters.}
\label{times_5x5}
\scriptsize
\begin{tabular*}{\columnwidth}{@{}@{\extracolsep{\fill}}llrr@{}}
\hline
\textbf{Algorithm} & \textbf{GPU kernel} & \multicolumn{1}{c}{\textbf{A}} & \multicolumn{1}{c}{\textbf{B}} \\\hline
\textbf{Winograd} & winogradForwardData4x4 & 13.82 & 13.89 \\
\textbf{non-fused} & winogradForwardFilter4x4 & 9.15 & 9.73 \\
 & volta\_sgemm\_128x64\_nn & 34.91 & 35.36 \\
 & winogradForwardOutput4x4 & 16.92 & 17.60 \\
 & \textit{\textbf{Total}} & \textit{74.80} & \textit{\textbf{76.58}} \\
 &  &  &  \\
\textbf{Our} & scalar\_prods\_kernel & 16.80 & 107.58 \\
\textbf{impl.} & sum\_kernel & 5.70 & 9.02 \\
 & \textit{\textbf{Total}} & \textit{\textbf{22.50}} & \textit{116.60} \\
\hline
\tablenotetwo{\rule{0pt}{2.5ex}\textbf{A}: 7-1-5-128-48, \textbf{B}: 7-8-5-128-48.}
\end{tabular*}
\end{table} 

\section{Related Work} \label{sect_relatedwork}

The main idea behind the GEMM-based convolution approach \cite{Chetlur_cuDNN14} is to convert a convolution into a matrix--matrix multiplication, thus being able to exploit already existing high-performance GEMM implementations. However, the data transformations required in this process may be too costly for naive implementations to be competitive (e.g. the best GEMM variant from cuDNN is usually the implicit with offsets precomputation).

Several optimization techniques for CNNs focus on algorithmic modifications to reduce the number of operations (mainly multiplications). As we already introduced in Section~\ref{sect_background}, Winograd--based~\cite{Lavin15b_winograd} and FFT--based~\cite{VasilacheJMCPL14_fft} convolutions are two examples of this approach. In both cases, inputs and filter are multiplied after being transformed into another domain, obtaining the convolution output employing less multiplication operations. However, considering only the theoretical number of multiplications fails to account for the possible overheads of performing the transformations. These overheads are most noticeable for convolutions with lower computational load, and when the transformations are implemented as separate GPU kernels like in the non-fused variant present in cuDNN.

Another proposal within this category is to reduce the arithmetic
complexity eliminating the redundant neuron connections
\cite{sparse_Han15}. For this, authors in that paper perform an initial training to assess which weights are irrelevant. Afterwards, they prune them by setting to zero the weights below a certain threshold. This converts the filters into sparse volumes, which combined with the dynamic sparsity of inputs from ReLU activations, enables them to reduce the number of multiplications needed for the convolution.

A recent paper by Xingyu Liu {\it et al.} \cite{sparse_wino_Liu18} combines the Winograd convolution with the pruning technique. They reorder some operations to be able to keep inputs and filters sparse in the Winograd domain, and thus reduce the cost of the point-wise multiplication. For the inputs, they apply the ReLU activation function of the previous layer after the inputs are already transformed to the Winograd domain. The filters are trained in the Winograd domain, avoiding the need for a domain transformation. To obtain sparse filters, they apply the pruning directly to the Winograd-domain filters. Another work by Jongsoo Park {\it et al.} \cite{iclr_ParkLWT0CD17} also uses pruning to improve the performance of convolution operations in CNNs. They present a dense-matrix-with-sparse-matrix multiplication implementation for Intel Atom, Xeon and Xeon Phy that enables the use of pruned filters in convolutional layers, thus reducing the storage and computation cost of these layers.

A different approach is presented by Chao Li {\it et al.}~\cite{mem_layout_Li16}. Instead of focusing on the computational efficiency, they focus on the memory efficiency of GPU implementations of convolution. As part of their work, they study how the tensors' data-layout affect the neural network performance, and propose some optimizations. Similar to this work, our convolution design pays special attention to the memory efficiency.

In a wider scope, there are several works that present other implementations of convolution operations to improve specific scenarios of CNNs training and deployment. For example, the work presented by Nicoli Dryden {\it et al.} \cite{cnn_dist_train_dryden19} focuses on improving the performance of distributed CNN training, and as part of their effort, they present a convolution operation implementation that is better suited to exploit model-parallelism.

\section{Conclusions and Future Work} \label{sect_concl_future}

After introducing the main features of the convolution operation used in CNNs, we presented our design for an implementation of convolution for GPUs. As part of this presentation, we thoroughly discussed the trade-offs involved in such design, to finally settle with an implementation that maximizes the exploitation of the data reuse available in the convolution operation, and performs efficient memory accesses without the need for data-layout transformations by selecting the most appropriate tensor layout. We evaluated our implementation with more than 600 convolution parameter configurations from the forward propagation phase of five of the most well known CNNs. We compared the performance of our convolution with all of the state--of--the--art convolution variants available in cuDNN, the reference library of GPU operations for deep learning. The results show that our implementation is highly competitive for certain parameter intervals, especially convolutions with 1$\times$1 filters and small batch size, where we obtain speedups of up to 2.29$\times$, which will contribute to the overall performance improvement without introducing any drawbacks due to the automatic convolution algorithm selection present in current machine learning frameworks.

As future work, we plan to explore the viability of a more sophisticated logic to distribute the work among GPU threads and thread blocks. The execution profiling suggests that, for some configurations, we could apply a work-fusion optimization to improve our design. For example, fusing the work of several thread blocks could be a potential optimization for configurations with large number of filters and small to medium depth. However, finding the optimal work distribution for all configurations is not straightforward, because it depends on the specific combination of the five convolution parameters.


%
%

\bibliographystyle{spmpsci}      
\bibliography{biblio}   

%
%

\end{document}